\documentclass[sn-mathphys-num]{sn-jnl}%

\usepackage{amssymb}
\usepackage{amsmath}

\usepackage{lineno}
\usepackage{eurosym}
\usepackage[ruled,vlined]{algorithm2e}
\usepackage{subfig}
\usepackage{url}
\usepackage{xcolor}

\usepackage{natbib}
\usepackage{graphicx}%
\usepackage{multirow}%
\usepackage{amsmath,amssymb,amsfonts}%
\usepackage{amsthm}%
\usepackage{mathrsfs}%
\usepackage[title]{appendix}%
\usepackage{xcolor}%
\usepackage{textcomp}%
\usepackage{manyfoot}%
\usepackage{booktabs}%
\usepackage{listings}%

\newcommand{\mathsc}[1]{{\normalfont\textsc{#1}}}

\newcommand{\partitions}{p}
\newcommand{\partitionsPerDim}{m}
\newcommand{\cores}{c}

\newcommand{\sky}{\mathsc{Sky}}

\newcommand{\sortedRep}{\mathsc{Sorted}\xspace}
\newcommand{\regionRep}{\mathsc{Region}\xspace}
\newcommand{\noseq}{\mathsc{NoSeq}\xspace}
\newcommand{\random}{\mathsc{Random}\xspace}
\newcommand{\grid}{\mathsc{Grid}\xspace}
\newcommand{\angular}{\mathsc{Angular}\xspace}
\newcommand{\sliced}{\mathsc{Sliced}\xspace}
\newcommand{\slicedrep}{\mathsc{Sliced+}\xspace}
\newcommand{\angularrep}{\mathsc{Angular+}\xspace}

\newcommand{\dominates}{\prec}

\newcommand{\dimensions}{d}
\newcommand{\monotoneFunctions}{\mathtt{MF}}

\newcommand{\positivereals}{\mathbb{R^+}}

\newcommand{\dominanceRegion}{\mathsc{DR}}

\newcommand{\size}{N}

\newcommand{\anticorrelated}{\texttt{ANT}\xspace}

\newcommand{\household}{\texttt{HOU}\xspace}

\newcommand{\zillow}{\texttt{RES}\xspace}

\newcommand{\griddominates}{\prec_G}
\newcommand{\weaklygriddominates}{\preceq_G}

\newcommand{\mycomment}[1]{}

\newtheorem{theorem}{Theorem}
\newtheorem{proposition}{Proposition}
\newtheorem{definition}{Definition}
\newtheorem{example}[theorem]{Example}

\newcommand{\logSep}{\,.\,\,}

\def\codeif{\mbox{\upshape\textbf{if}}}
\def\codethen{\mbox{\upshape\textbf{then}}}

\def\codeforeach{\mbox{\upshape\textbf{for each}}}
\def\codedo{\mbox{\upshape\textbf{do}}}

\def\codereturn{\mbox{\upshape\textbf{return}}}

\def\codecontinue{\mbox{\upshape\textbf{continue}}}

\def\codeparallel{\mbox{\upshape\textbf{parallel}}}

\begin{document}

\title{Optimization Strategies for Parallel Computation of Skylines}

\author[1]{\fnm{Paolo} \sur{Ciaccia}}\email{paolo.ciaccia@unibo.it}

\author*[2]{\fnm{Davide} \sur{Martinenghi}}\email{davide.martinenghi@polimi.it}

\affil[1]{\orgdiv{DISI}, \orgname{Universit\`a di Bologna}, \orgaddress{\street{Viale Risorgimento 2}, \city{Bologna}, \postcode{40136}, \country{Italy}}}

\affil*[2]{\orgdiv{DEIB}, \orgname{Politecnico di Milano}, \orgaddress{\street{Piazza Leonardo 32}, \city{Milan}, \postcode{20133}, \country{Italy}}}

\abstract{
Skyline queries are one of the most widely adopted tools for Multi-Criteria Analysis, with applications covering diverse domains, including, e.g., Database Systems, Data Mining, and Decision Making. Skylines indeed offer a useful overview of the most suitable alternatives in a dataset, while discarding all the options that are dominated by (i.e., worse than) others.

The intrinsically quadratic complexity associated with skyline computation has pushed researchers to identify strategies for parallelizing the task, particularly by partitioning the dataset at hand.
In this paper, after reviewing the main partitioning approaches available in the relevant literature, we propose two orthogonal optimization strategies for reducing the computational overhead, and compare them experimentally in a multi-core environment equipped with PySpark.
}

\keywords{skyline, parallel computation, PySpark}

\maketitle

\section{Introduction}
\label{intro}

One of the goals of Multi-Criteria Analysis is to find the most interesting alternatives available in a dataset whose elements are described by several attributes. This problem occurs very frequently in a large variety of data-intensive application contexts, including Database Systems, Data Mining, and Decision Making. The advent of Big Data scenarios has made the importance of this problem even more central, since users and companies need to search within larger and larger datasets.

To this end, skyline queries are one of the most widely adopted tools, allowing their users to retain only the alternatives that are not dominated by others. Here, alternative $a$ is said to dominate $b$ if, for all attributes, $a$ is never worse than $b$, and strictly better for at least one attribute. As is well-known, each non-dominated alternative is guaranteed to be the top-$1$ choice for at least one (monotone) function $f$ of the attributes, i.e., the first one if we rank the dataset according to $f$. This means that, if we associate every possible user $u$ with the function $f_u$ they would ideally use for ranking the alternatives, the skyline (i.e., the set of non-dominated alternatives) contains exactly those choices that are top-$1$ for at least one user, thereby offering a useful overview of all the potentially optimal elements of a dataset.

Skyline queries are also known to generally exhibit an asymptotic quadratic complexity in the number of elements of the dataset, which makes them a much less desirable solution in Big Data contexts.
This has led the research community to investigate, on one hand, new algorithms incurring lower computational overhead (although, apart from cases of very low dimensionality of the dataset~\cite{DBLP:journals/jacm/KungLP75}, no sub-quadratic solutions are possible) and, on the other hand, strategies for partitioning the dataset so as to process each partition in parallel to discard most of the dominated alternatives and, thus, reduce the overall computation time.
Such partitioning schemes can be effectively applied by resorting to parallelized frameworks such as PySpark, where the different ``compute nodes'' (e.g., the different cores available in a multi-core setting) process data independently and synchronize with each other at specific moments determined by the framework. 

The overall idea of existing approaches is to determine the skyline via a 2-phase process. First, one obtains from each partition a so-called ``local skyline'' (i.e., the non-dominated alternatives present in that partition); second, once all the local skylines have been computed, they can be merged to form a pruned dataset, on top of which a final round of (sequential) skyline computation phase is applied.
The common objective of existing approaches is therefore to partition the data so as to prune as many dominated alternative as possible already in the local skyline computation phase, so that the last phase will have to handle a minimally sized dataset.

In this paper we propose two novel optimization strategies that are \emph{orthogonal} to the adopted partitioning criterion, and as such have general applicability. The first strategy applies to the first phase of the above-described schema, and consists in passing some ``good'' alternatives across the partitions, thus aiming to reduce the size of the dataset to deal with in the second phase. The second strategy builds on the observation that in the approach adopted by previous works, the second phase completely gives up with parallelism and then becomes the real bottleneck of the whole process. To obviate this limitation, we show that \emph{even the second phase can be fully parallelized}, a fact that has been mostly overlooked in the past.

Overall, the contributions of this work can be summarized as follows.
First, we provide a survey of the main partitioning strategies available in the relevant literature.
Second, after discussing the overall scheme for parallelizing skyline computation on a framework such as PySpark, we discuss orthogonal optimization layers, particularly proposing the novel technique of representative filtering.
Third, we compare the resulting combinations of (sequential) algorithms and partitioning strategies with one another from an experimental point of view, by considering a number of parameters regarding the partitioning strategy, the dataset, and the execution environment.

\section{Preliminaries}
\label{sec:prelim}

Since Multi-Criteria Analysis generally focuses on numeric attributes, in the following we shall refer to datasets consisting of attributes whose domain is, without loss of generality, the set of non-negative real numbers $\positivereals$. Under this assumption, and consistently with standard database terminology, a schema $S$ is a set of attributes $\{A_1,\ldots, A_\dimensions\}$, and a tuple $t=\langle v_1,\ldots,v_\dimensions\rangle$ over $S$ is a function associating each attribute $A_i\in S$ with a value $v_i$, also denoted $t[A_i]$, in $\positivereals$; a relation over $S$ is a set of tuples over $S$.

The notion of \emph{skyline}, introduced in~\cite{DBLP:conf/icde/BorzsonyiKS01}, is based on the concept of \emph{dominance} between tuples: a skyline query takes a relation $r$ as input and returns the set tuples in $r$ that are dominated by no other tuples in $r$.

\begin{definition}
Let $S$ be a schema and $t$ and $s$ two tuples over $S$. We say that $t$ \emph{dominates} $s$, denoted $t\dominates s$, if, for every attribute $A\in S$, $t[A]\leq s[A]$ holds and there exists an attribute $A' \in S$ such that $t[A']<s[A']$ holds.
The \emph{skyline} $\sky(r)$ of a relation $r$ over $S$ is the set $\{t \in r \mid \nexists s\in r \logSep s\dominates t\}$.
\end{definition}

Although in this paper we shall consider attributes such as ``cost'' and ``distance'', for which smaller values are preferable, the opposite convention would of course also be possible.
We observe that tuples can be regarded as points in $\positivereals^\dimensions$; as such, given a tuple $t$, we introduce the notion of dominance region of $t$ as the set of all points corresponding to tuples that would be dominated by $t$.
\begin{definition}
Let $t$ be a tuple over a schema $S$ with $\dimensions$ attributes. The \emph{dominance region} of $t$ is the set $\dominanceRegion(t)=\{s\in\positivereals^\dimensions \mid t\dominates s\}$.
\end{definition}

It is customary to associate tuples with numeric scores. This is done by resorting to a so-called \emph{scoring function}, which is most typically a monotone function (e.g., a weighted sum) applied to the tuple's attribute values. For a schema $S=\{A_1,\ldots, A_\dimensions\}$ and a tuple $t$ over $S$, a scoring function $f$ returns a score
$f(t[A_1],\ldots, t[A_\dimensions])\in \positivereals$,
also indicated $f(t)$ for brevity;
$f$ is \emph{monotone} if, for all tuples $t$ and $s$ over $S$, we have
$(\forall A\in S \logSep t[A]\leq s[A])\rightarrow f(t) \leq f(s)$.

According to our convention for attribute values, lower scores are preferred to higher ones.
It is interesting to observe that, for every tuple $t\in\sky(r)$, there exists a monotone scoring function $f$ such that $f(t)$ is the minimum (i.e., optimal) score for all tuples in $r$.
Under this perspective, skylines can equivalently be viewed as the set of optimal tuples according to at least one scoring function~\cite{DBLP:journals/sigmod/ChomickiCM13}, i.e., $\sky(r)=\{t\in r \mid \exists f \in \monotoneFunctions \logSep \forall s \in r \logSep s \neq t \rightarrow f(t) < f(s)\}$, where $\monotoneFunctions$ denotes the set of all monotone functions.

\begin{figure}[t]%
\centering%
\includegraphics[width=0.5\textwidth]{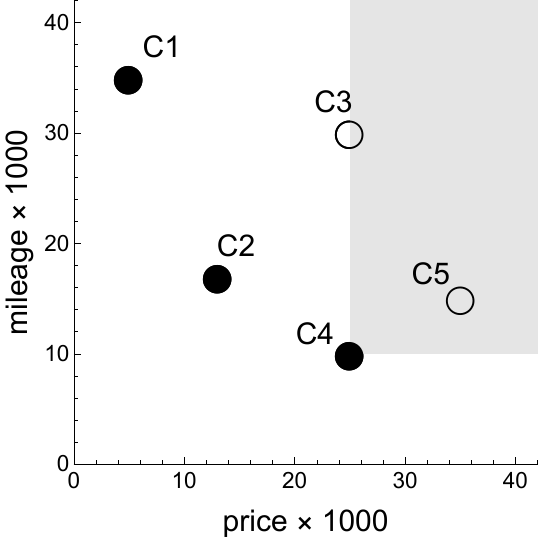}
\caption{An example dataset of used cars. The filled disks represent tuples in the skyline. The gray area is the dominance region of car C4.}\label{fig:cars-intro}
\end{figure}

\begin{example}
To fix these preliminary notions, consider the dataset shown in Figure~\ref{fig:cars-intro},
reporting mileage (in Km$\cdot 10^3$) and price (in K\euro) of used cars. Clearly, $C_3$ and $C_5$ are dominated by $C_4$ (whose dominance region is shown in gray), while all the other cars are non-dominated. Therefore, the skyline is $\{C_1, C_2, C_4\}$. Car $C_1$ is the top-1 if we rank all cars by price (e.g., via a scoring function $f_1(x,y)=x$), $C_4$ is the best choice by mileage (e.g., via $f_2(x,y)=y$), while $C_2$ is the best option if we give the same importance to price and mileage (e.g., via $f_3(x,y)=x+y$).
\end{example}

Several algorithms are available in the literature for computing the skyline of a dataset in a traditional, non-parallel setting. Any of these algorithms can be used in a parallel framework for computing the local skylines at the processing nodes.
Without loss of generality, in this paper we adopt the Sort Filter Skyline (SFS) algorithm~\cite{DBLP:conf/icde/ChomickiGGL03} (other algorithms will be discussed in Section~\ref{sec:related}). 

\begin{algorithm}[t]
\scalebox{.95}
   {

    \begin{minipage}{1.33\textwidth}
	\begin{enumerate}
	    \item[Input:] \emph{relation $r$, monotone function $f$ over $r$'s attributes}
	    \item[Output:] $\sky(r)$ 
		\item $w := \emptyset$  // \textit{the window, initially empty}
		\item $m := $ sorted version of $r$ according to $f$
	    \item\label{line:outer-for-sfs} \codeforeach\ $t$ in $m$ \codedo
	    \item\label{line:dominance-check-sfs} \quad \codeforeach\ $u$ in $w$ \codedo
	    \item \quad \quad  \codeif\ $u\dominates t$ \codethen\ \codecontinue\ to line~\ref{line:outer-for-sfs}
	    \item\label{line:window-grows-sfs} \quad $w := w \cup \{t\}$
	    \item \codereturn\ $w$
	\end{enumerate}	    
	\end{minipage}
   }
	\caption{The SFS algorithm.}
	\label{alg:sfs}
\end{algorithm}

The SFS algorithm, whose pseudo-code is shown in Algorithm~\ref{alg:sfs}, considers a pre-processing phase that sorts the tuples according to a monotone function of their attributes. The ordered dataset thus becomes a \emph{topological sort} with respect to dominance, i.e., if tuple $s$ follows tuple $t$ in the ordering, then $s$ cannot dominate $t$.
Subsequently, SFS scans the dataset ($r$) while maintaining a ``window'' ($w$) in main memory containing the candidate skyline tuples. Each tuple $t$ read from the dataset is checked for dominance against all the tuples in $w$. If $t$ is dominated, then it is definitively discarded, otherwise it is added to $w$; notice that, due to the topological sort property, $t$ cannot dominate any tuple in the window $w$. At the end of the process, $w$ will contain the skyline.

\section{Parallel Algorithms}
\label{sec:parallel}

We now present a general scheme for parallelizing the computation of the skyline, based on the idea that, if we partition the dataset, each partition can be processed independently and in parallel so as to produce a ``local'' skyline. The union of all the local skylines will generally be much smaller than the original dataset and the (global) skyline can finally be obtained by applying a last (sequential) round of removal of dominated tuples.

The correctness of the described approach is guaranteed by the following property of a relation $r$ partitioned into sub-relations $r_1,\ldots, r_\partitions$.
\begin{proposition}
Let $r=r_1 \cup \ldots \cup r_\partitions$, with $r_i\cap r_j=\emptyset$ for $i\neq j$. Then $\sky(r)=\sky(\sky(r_1)\cup\ldots\cup\sky(r_\partitions))$.
\end{proposition}

\begin{algorithm}[t]
\scalebox{.95}
   {
    \begin{minipage}{1.33\textwidth}
	\begin{enumerate}
	    \item[Input:] \emph{relation $r$, number of partitions $\partitions$}
	    \item[Output:] $\sky(r)$ 
		\item $ls := \emptyset$  // \textit{the local skyline}
		\item\label{line:partitioning-pattern} $(r_1,\ldots,r_\partitions, meta) := \texttt{Partition}(r,\partitions)$  // \textit{partitions and meta-information}
	    \item\label{line:parallel-for} \codeparallel\ \codeforeach\ $r_i$ in $r_1,\ldots,r_\partitions$ \codedo
	    \item\label{line:union-of-local-skylines} \quad $ls := ls \cup \texttt{ComputeLocalSet}(r_i,meta)$
	    \item\label{line:last-round-pattern} \codereturn\ $\sky(ls)$
	\end{enumerate}	    
	\end{minipage}
   }
	\caption{Algorithmic pattern for parallel skyline computation.}
	\label{alg:parallel-scheme}
\end{algorithm}

The algorithmic pattern shown in Algorithm~\ref{alg:parallel-scheme} consists of three main phases. The first one is the partitioning of $r$ into $r_1,\ldots,r_\partitions$ (line~\ref{line:partitioning-pattern}); as a by-product of this phase, meta-information that might be useful for speeding up the second phase may be collected.
The second phase consists of computing the (local) skyline independently on each partition $r_i$ (line~\ref{line:parallel-for}), by exploiting as much as possible any opportunity for parallelization supported by the execution framework, as indicated by the $\codeparallel$ keyword, as well as meta-information passed from the previous phase, which may be used to speed up the computation of local skylines or to orchestrate the general process.
During the last phase, all the local skylines are merged into a single set (line~\ref{line:union-of-local-skylines}) and any remaining dominated tuples are removed with a final, sequential round (line~\ref{line:last-round-pattern}).

The specific partitioning strategy, along with the availability of resources for parallelization, may heavily affect the execution time.

In the following subsections, we are going to describe four partitioning strategies -- three are taken from the literature and one is novel. 
Although all of them show potential for speeding up execution time compared to the sequential version, thanks to work parallelization, some partitioning strategies prove to be more effective than others, as we will see through experiments in Section~\ref{sec:experiments}.

\begin{figure}%
\centering
\subfloat[][{\random}]
{\includegraphics[width=0.22\textwidth]{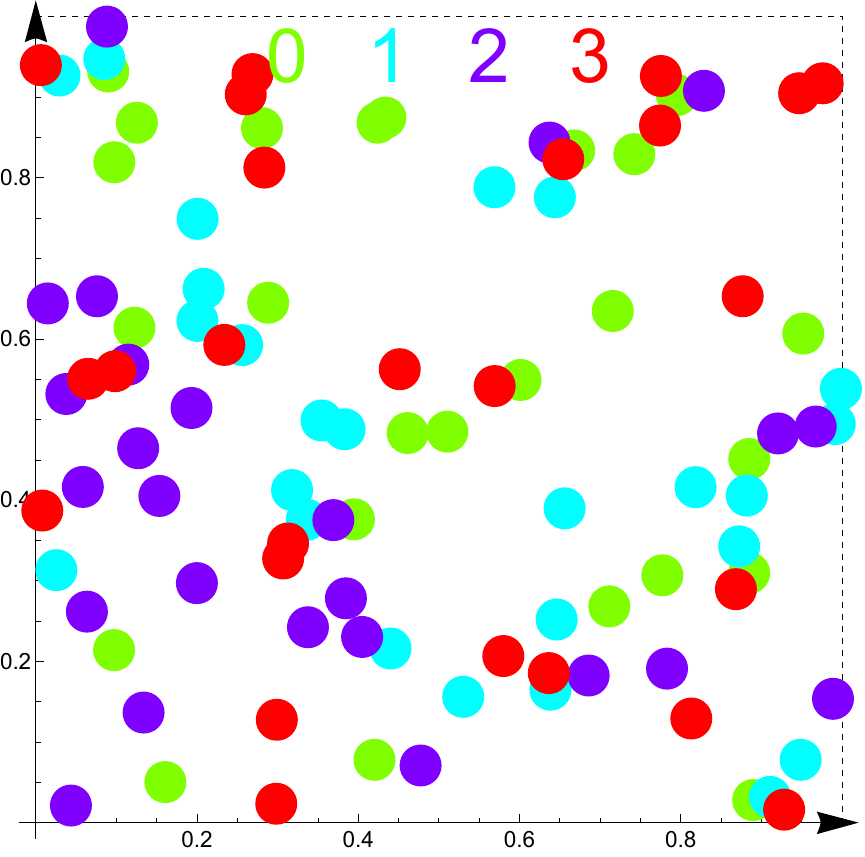}\label{fig:randomPartitioning}}%
\quad
\subfloat[][{\grid}]
{\includegraphics[width=0.22\textwidth]{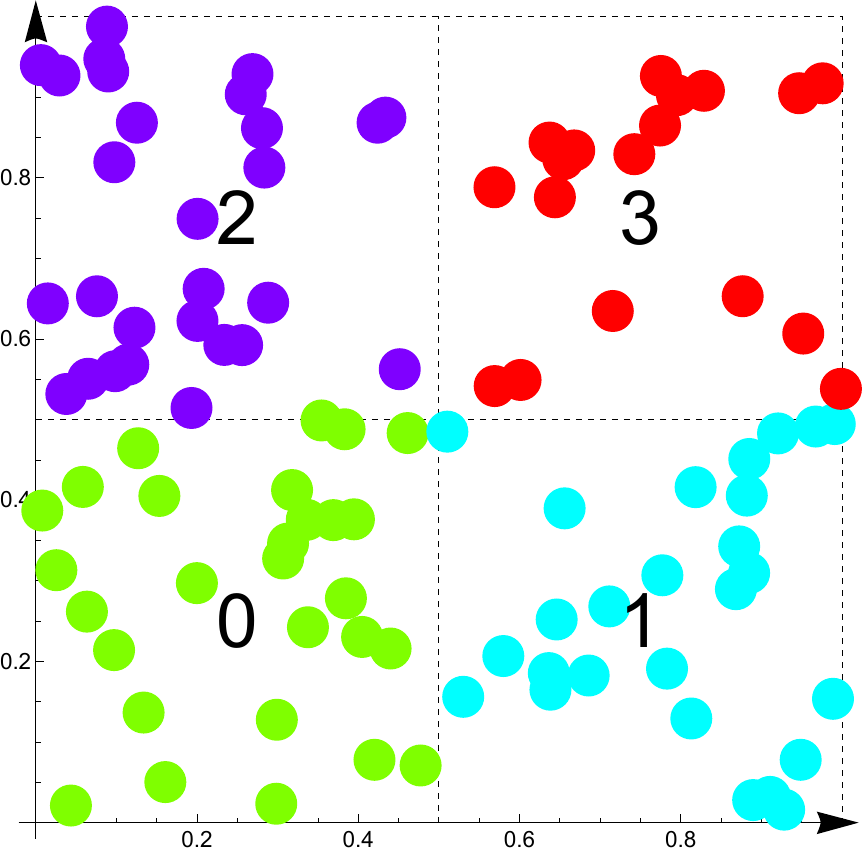}\label{fig:gridPartitioning}}%
\quad
\subfloat[][{\angular}]
{\includegraphics[width=0.22\textwidth]{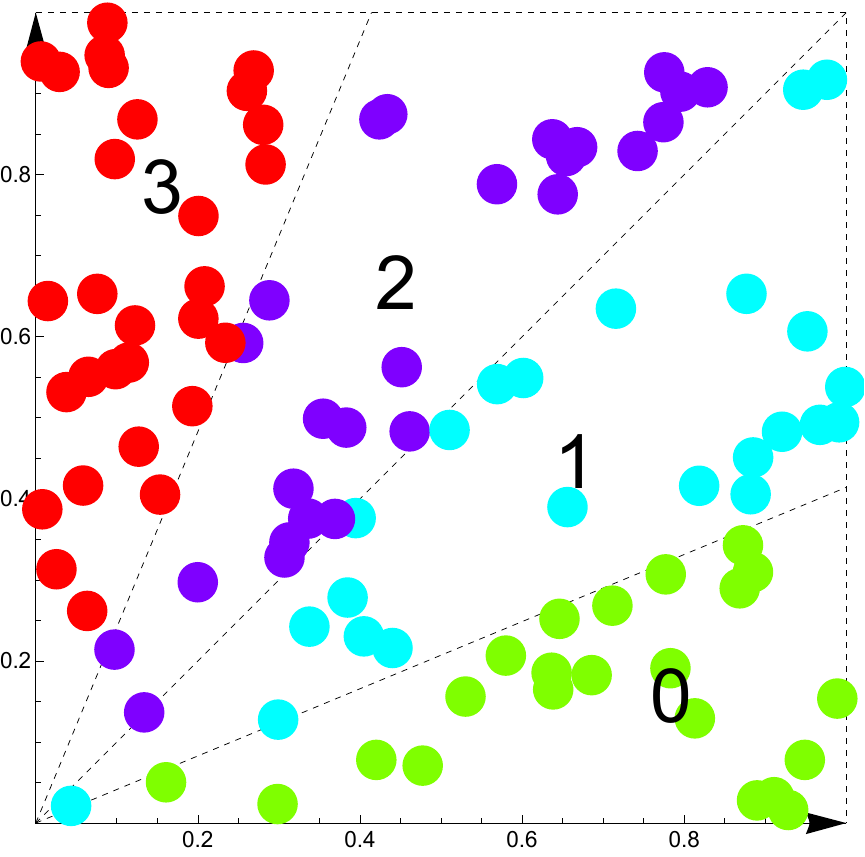}\label{fig:anglePartitioning}}%
\quad
\subfloat[][{\sliced}]
{\includegraphics[width=0.22\textwidth]{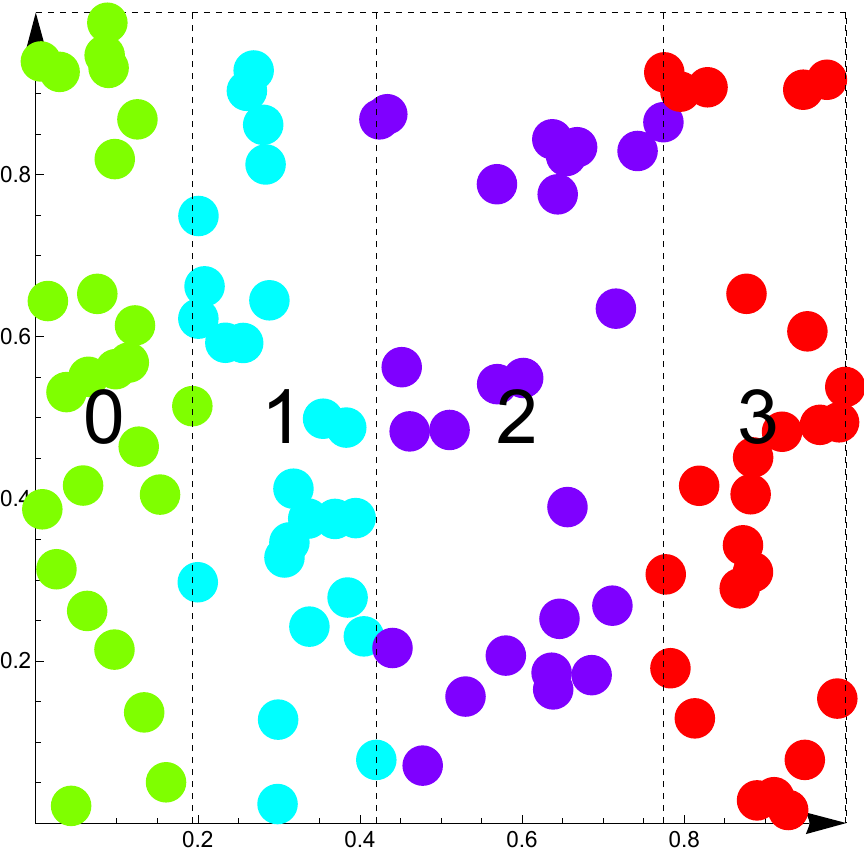}\label{fig:slicedPartitioning}}%
\\
\subfloat[][{\random}]
{\includegraphics[width=0.22\textwidth]{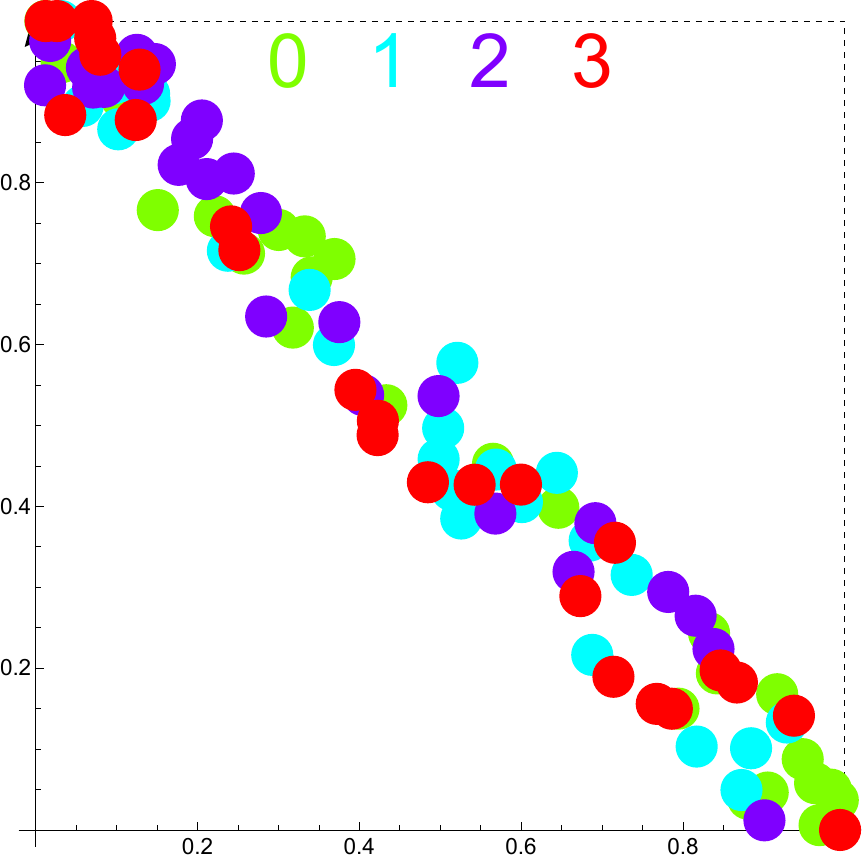}\label{fig:randomPartitioning-anticorrelated}}%
\quad
\subfloat[][{\grid}]
{\includegraphics[width=0.22\textwidth]{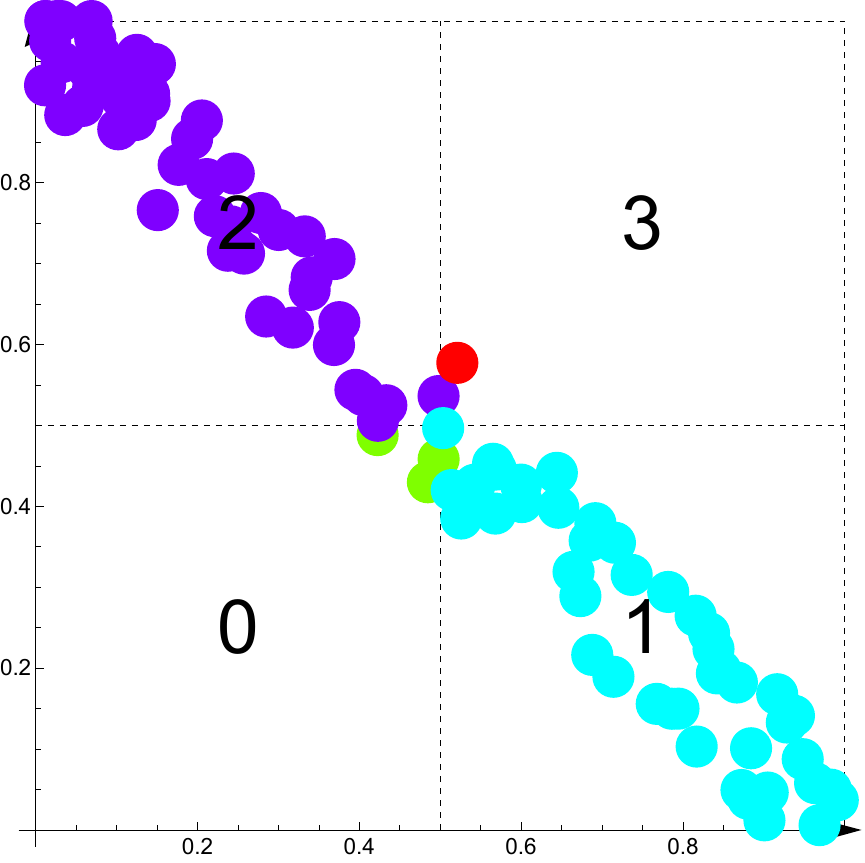}\label{fig:gridPartitioning-anticorrelated}}%
\quad
\subfloat[][{\angular}]
{\includegraphics[width=0.22\textwidth]{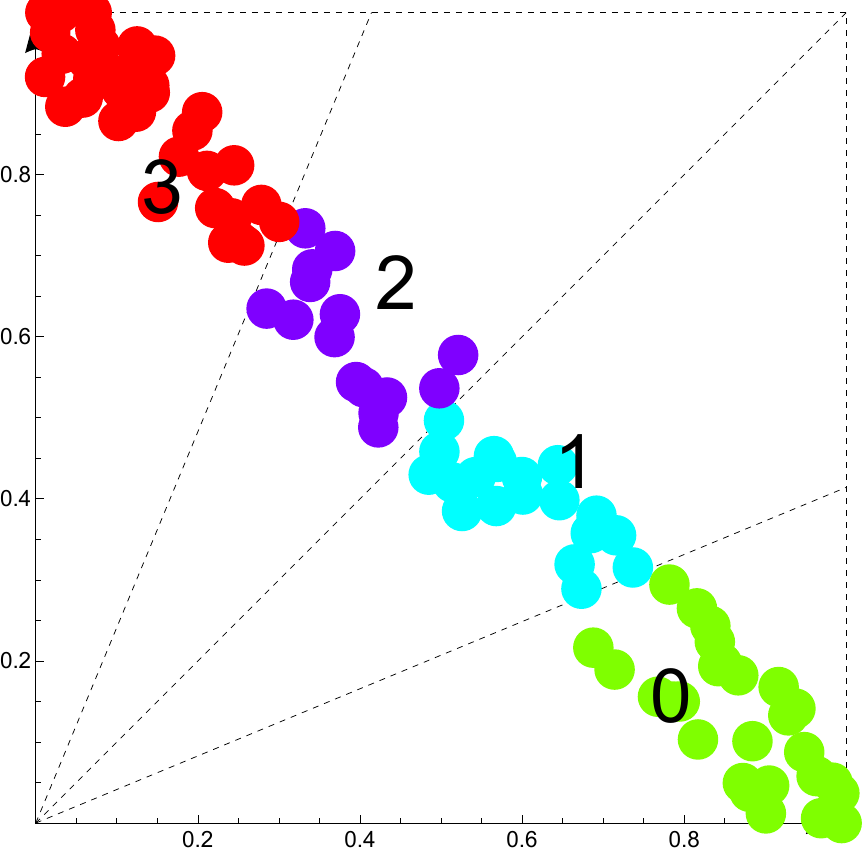}\label{fig:anglePartitioning-anticorrelated}}%
\quad
\subfloat[][{\sliced}]
{\includegraphics[width=0.22\textwidth]{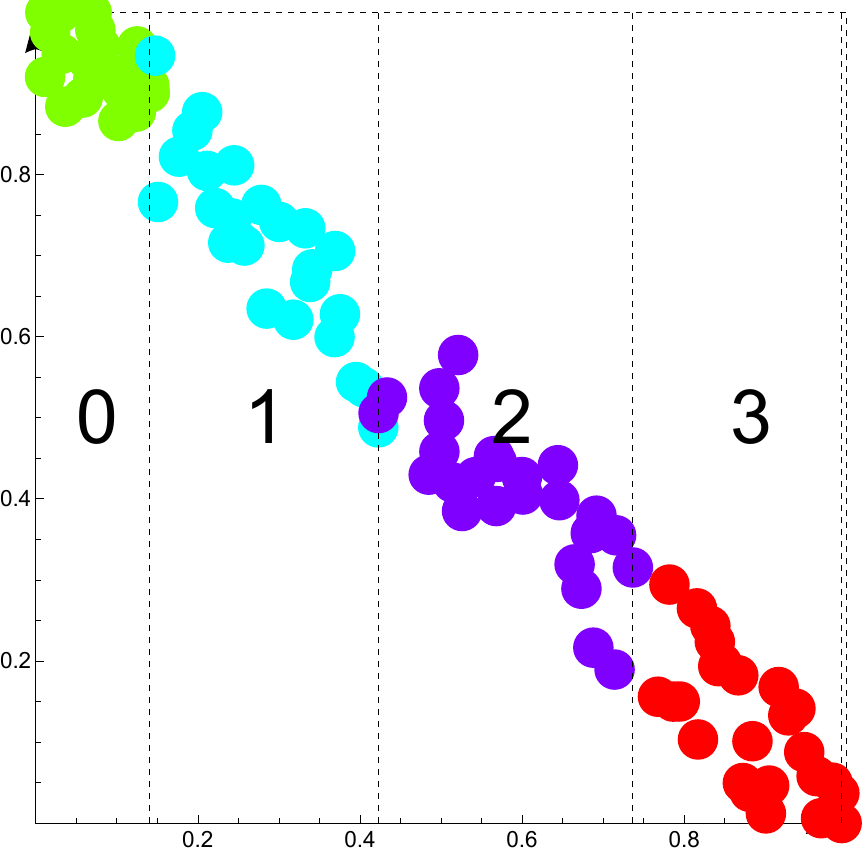}\label{fig:slicedPartitioning-anticorrelated}}%
\caption{Partitioning Strategies. First row: uniform dataset. Second row: anticorrelated dataset.}\label{fig:precision-recall}%
\end{figure}

\subsection{Random Partitioning}

A partitioning strategy typically used as a baseline is Random Partitioning~\cite{DBLP:conf/hpcs/Cosgaya-LozanoRZ07}, henceforth simply indicated as \random. The aim of \random is to ensure that each partition represents a sample of the original dataset with similar structural characteristics.
To achieve this, tuples are randomly distributed across the various partitions and all partitions are equi-numerous.
The simplicity of this method eliminates the need for any pre-processing of the data. However, it produces comparatively large local skylines with respect to other partitioning strategies. As a consequence of this, the final sequential round of Algorithm~\ref{alg:parallel-scheme} will be more costly.
Figure~\ref{fig:randomPartitioning} shows a possible partitioning of a uniformly distributed dataset of 100 tuples to 4 partitions, where different partitions are represented with different colors. Its effects on anti-correlated data are shown in Figure~\ref{fig:randomPartitioning-anticorrelated}.

\subsection{Grid Space Partitioning}

Grid Partitioning~\cite{DBLP:conf/edbt/MullesgaardPLZ14} (\grid) divides the space into a grid of equally sized cells, each of which corresponds to a different partition.
Each dimension is divided into $\partitionsPerDim$ parts, resulting in a total of $\partitions=\partitionsPerDim^\dimensions$ partitions, where $\dimensions$ denotes the total number of dimensions.
Besides assigning tuples to partitions, with this strategy we can also leverage a sort of dominance relationship that applies to grid cells (and thus to partitions) allowing us to avoid processing certain partitions completely (a technique called \emph{Grid Filtering}).

In particular, for a given cell $c_i$ (corresponding to partition $r_i$), let us consider its \emph{grid coordinates}, $\langle c_i[1],\ldots,c_i[\dimensions]\rangle$, with $1\leq c_i[j]\leq m$, $j=1,\ldots,\dimensions$.

\begin{definition}[Grid dominance]
Given two grid cells $c_i$ and $c_h$ we say that $c_i$ \emph{grid-dominates} $c_h$, denoted $c_i \griddominates c_h$, if for every dimension $j$, $j=1,\ldots,\dimensions$, we have $c_i[j] < c_h[j]$.  
\end{definition}

If $c_i$ grid-dominates $c_h$ then all tuples in $r_i$ dominate all tuples in $r_j$ and, therefore, if $r_i$ contains at least one tuple, partition $r_j$ can be disregarded.

Assuming, without loss of generality, that all values are normalized in the $[0,1]$ interval, a simple index $\partitions(t)$ can be computed as follows to identify the partition for each tuple $t$ in the dataset:

\[
\partitions(t) = \sum_{i=1}^{\dimensions} \lfloor t[A_i]\cdot \partitionsPerDim \rfloor \cdot \partitionsPerDim^{i-1}
\]
\noindent where $A_i$ is the $i$-th attribute and $\partitionsPerDim$ is the number of slices in which each of the $\dimensions$ dimensions is divided.

Figure~\ref{fig:gridPartitioning} shows the same dataset as Figure~\ref{fig:randomPartitioning} and a grid partitioning with $\partitionsPerDim=2$ partitions per dimension, amounting to a total of $4$ partitions (grid cells).
As we shall see, using Algorithm~\ref{alg:parallel-scheme} on top of \grid will cause to retain many tuples in the local skylines that will later be discarded from the global skyline, thus incurring unneeded computational overhead.
Grid filtering can be used to mitigate this effect.
Unlike \random, \grid gives no control on the number of tuples of each partition, which may end up being extremely unbalanced, as Figure~\ref{fig:gridPartitioning-anticorrelated} shows in the case of anti-correlated data. 
There is also limited control over the total number of partitions, since they depend on the number of dimensions $\dimensions$ and the number of slices per dimension $\partitionsPerDim$. Still, compared to \random, \grid manages to eliminate more tuples during local skyline computation. 

\subsection{Angle-based Space Partitioning}

The main idea behind Angle-based Partitioning~\cite{DBLP:conf/sigmod/VlachouDK08} (henceforth: \angular) is to partition the space based on angular coordinates, after converting Cartesian to hyper-spherical coordinates.
The main benefits include a better workload balance, as each partition includes both good and bad tuples, leading to local skylines with fewer globally dominated tuples, thus reducing the workload for the sequential phase compared to \grid and \random.
Unlike \grid, \angular does not allow for pruning. However, in the next section we will present a filtering technique addressing this limitation.

Here, too, we map every tuple $t$ to an index $\partitions(t)$ corresponding to the partition for $t$.
The index is computed based on hyper-spherical coordinates, including a radial coordinate $r$ and $\dimensions-1$ angular coordinates $\varphi_1, \ldots, \varphi_{\dimensions-1}$. This transformation from Cartesian coordinates $\langle x_1,\ldots, x_\dimensions\rangle$ is as follows:
\[
r=\sqrt{\sum_{j=1}^\dimensions x_j^2};
\quad
\tan(\varphi_i)=\frac{\sqrt{\sum_{j=i+1}^\dimensions x_j^2}}{x_i} \mbox{ for } 1\leq i \leq \dimensions-1,
\]
where $0\leq \varphi_i \leq \frac{\pi}{2}$ for all $i$, as in~\cite{DBLP:conf/sigmod/VlachouDK08}.

The index of tuple $t=\langle x_1,\ldots, x_\dimensions\rangle$ is then computed as follows:

\begin{equation}\label{eq:index-angular}
\partitions(t)=\sum_{i=1}^{\dimensions-1}\left\lfloor \frac{2\varphi_i}{\pi}\partitionsPerDim \right\rfloor
\partitionsPerDim^{i-1}
\end{equation}
where $\partitionsPerDim$ is, again, the number of slices in which each (angular) dimension is divided, which essentially amounts to grid partitioning on angular coordinates.

Figure~\ref{fig:anglePartitioning} shows \angular at work on a uniformly distributed dataset, and~\ref{fig:anglePartitioning-anticorrelated} on anti-correlated data, achieving a better workload balance than \grid.

\subsection{Sliced Partitioning (One-dimensional Slicing)}

The idea behind Sliced Partitioning (\sliced) is to sort the dataset with respect to one chosen dimension (possibly with the addition of a tie-breaking criterion to obtain a total ordering). Then, unlike \grid and \angular (when $\dimensions > 2$), any number of partitions with the same number of tuples can be easily obtained by scanning the sorted dataset.
The $i$-th tuple $t$ in the ordering is assigned to a partition, characterized as usual by an index $\partitions(t)$, in the following way:
\[
\partitions(t) = \left\lfloor \frac{(i-1)\cdot \partitions}{\size-1}\right\rfloor,
\]
where $\size$ is the number of tuples in the dataset.

Figures~\ref{fig:slicedPartitioning} and~\ref{fig:slicedPartitioning-anticorrelated} show the effect of \sliced on a uniform dataset and, respectively, an anti-correlated dataset, granting an assignment of the same number of tuples to each partition.

Additionally, we observe that, if the ordering is total, the subsequent local skyline computation phase can proceed, e.g., with SFS without presorting.

\section{Optimization Strategies for Parallel Algorithms}
\label{sec:improvements}

In this section, we describe two possible enhancements of the general pattern described in Algorithm~\ref{alg:parallel-scheme}.
The first enhancement consists of a filtering technique, termed \emph{Representative Filtering} and described in Section~\ref{sec:representative-filtering}, which aims to reduce the amount of tuples in the local skylines. The second one, described in Section~\ref{sec:all-parallel}, completely removes the final sequential phase by substituting it with a second parallelized phase.

\subsection{Representative Filtering}\label{sec:representative-filtering}

A promising idea to prune most of the globally dominated tuples already during local skyline computation is to pre-compute a few potentially ``strong'' tuples (the \emph{representatives}) from the entire dataset and share them as meta-information with the nodes taking care of each partition.
Any tuple dominated by a representative can be deleted with no further ado.
These representatives can be chosen according to several strategies, with a profound impact on the overall effectiveness of the algorithm.

There are various methods to select representative tuples. Clearly, the baseline strategy 
would be to randomly choose a fixed number of them either from the entire dataset or from each partition. The main disadvantage of this strategy is that it may include weak tuples with a low pruning power.

A more promising strategy (\sortedRep) consists in selecting the first tuples in each partition after sorting: by virtue of the topological sort property, they cannot be dominated by subsequent tuples and therefore are likely to dominate many of them.

Another strategy (\regionRep) consists in selecting the tuples with the largest associated dominance region, which can be interpreted as a measure of the potential of a tuple to dominate other tuples. Clearly, for this notion to be applicable, one needs to deal with finite regions, which can be achieved by simply normalizing the data in a finite range. For this reason, in the following we shall always implicitly refer to the $[0,1]$ domain for all attribues.
Under this assumption, the (hyper-)volume $V(\cdot)$ of the dominance region $\dominanceRegion(t)$ of a tuple $t$ can then be computed as
\[
V(\dominanceRegion(t))=\prod_{i=1}^\dimensions (1-t[i]).
\]

Independently of the strategy for choosing the representatives, any dominated tuples among these is discarded before sending them out as meta-information.
Then, if a point is not dominated by any representative, it is added to the set of non-dominated points.

The advantage of \sortedRep is that sorting is enough to obtain the representatives, while \regionRep requires an explicit step for computing them and does not work with non-normalized datasets.

\subsection{No sequential phase}\label{sec:all-parallel}
The final sequential phase is typically the most time-consuming in the execution of Algorithm~\ref{alg:parallel-scheme}. We now discuss how to eliminate the sequential phase completely.

Let $u_i = \sky(r_i)$, $i=1,\ldots,\partitions$, be the local skylines computed in the first phase, and let $u$ be their union. The key idea underlying our second optimization is to discard from each $u_i$ those tuples that are globally dominated (thus, not belonging to $\sky(r)$) by comparing $u_i$ with tuples in a proper subset of $u$, whose definition depends on the specific partitioning strategy. 

We start with the following:
\begin{definition}
The \emph{relative skyline} $\sky_c(r)$ of a relation $r$ \emph{with respect to a relation $c$} is defined as:
$$
\sky_c(r) = \{t \in r \mid \nexists s\in c \logSep s\dominates t\}.
$$
\end{definition}
Clearly, $\sky_r(r) = \sky(r)$ is the usual skyline.
Given above definition we can exploit the following identity to parallelize even the second phase.
\begin{proposition}\label{prop:noseq}
Let $r=r_1 \cup \ldots \cup r_\partitions$, with $r_i\cap r_j=\emptyset$ for $i\neq j$. 
Let $u_i = \sky(r_i)$, $u = u_1 \cup \ldots \cup u_\partitions$,
and $pd_i \subseteq u \setminus u_i$
be any subset such that the following holds $\forall t\in u_i$, $i=1,\ldots,\partitions$:
\begin{equation}\label{eq:potentialdominators}
t\not\in\sky(r) \implies \exists s\in pd_i\;\; \mbox{such that} \:\: s\dominates t.
\end{equation}
Then:
\begin{equation}\label{eq:noseq}
\sky(r)=\sky_{pd_1}(u_1)\cup\ldots\cup\sky_{pd_\partitions}(u_\partitions).
\end{equation}
\end{proposition}

Notice that $pd_i \subseteq u \setminus u_i$ (rather than $pd_i \subseteq u$) since it is meaningless to compare (again) tuples in $u_i$ with themselves. 
The $pd_i$ set contains the ``\emph{potential dominators}'' of tuples in $u_i$. For the \random and the \angular partitioning strategies, which do not generate partitions for which a dominance relationship can be established, the \emph{only} sound choice is to have $pd_i = u \setminus u_i$, since a tuple in $u_i$ can be dominated by tuples in any of the other partitions.

For the \grid partitioning method we can exploit a relaxed (weak) version of the grid-dominance relationsip among cells of the grid.
\begin{definition}[Weak grid dominance]
Given two grid cells $c_i$ and $c_h$ we say that $c_i$ \emph{weakly grid-dominates} $c_h$, denoted $c_i \weaklygriddominates c_h$, if  $c_i[j] \leq c_h[j]$, $j=1,\ldots,\dimensions$, and at least one inequality is not strict.
\end{definition}
Now, if $c_i$ \emph{does not} weakly grid-dominate $c_h$, this implies that no tuple in $u_i$ can dominate any tuple in $u_h$. This is sufficient to define
\[
pd_i = \{ u_j | c_j \weaklygriddominates c_i \} \quad i=1,\ldots,\partitions.
\]
Similar considerations apply to the \sliced partitioning strategy, for which it is immediate to derive:
\[
pd_i = \{ u_j | j < i \} \quad i=1,\ldots,\partitions.
\]

The above scheme, henceforth referred to as \noseq, also complies with Algorithm~\ref{alg:parallel-scheme}, where the computation of $\sky$ on line~\ref{line:last-round-pattern} uses, in turn, another instance of Algorithm~\ref{alg:parallel-scheme}, in which the meta-information passed to each parallel node is the union $u$ of the local skylines.
This approach is of course meaningful inasmuch as $u$ is smaller than the original dataset.

\section{Experiments}
\label{sec:experiments}

In this section, we test both the effectiveness and the efficiency of the proposed algorithmic pattern.
In particular, in Section~\ref{sec:effectiveness} we test the effectiveness of the filtering techniques that can be applied on top of Grid Partitioning (i.e., Grid Filtering) or generally with any partitioning scheme (i.e., Representative Filtering).

We measure several indicators of efficiency as discussed in Section~\ref{sec:efficiency} in different settings and various datasets of synthetic as well as real nature. Synthetic datasets are generated with a target number of tuples varying between 100K and 100M, a target number of dimensions between 2 and 7, and following a given distribution among uniform, correlated, and anticorrelated. The real datasets we consider are \household{}\footnote{\url{https://archive.ics.uci.edu/ dataset/235/individual+household+electric+power+consumption}} and \zillow{}\footnote{\url{https://www.zillow.com}}.
The former contains measurements of electricity consumption in a household with a sampling frequency of one minute for a period of almost 4 years, totaling 2,049,280 tuples and 7 dimensions after cleaning and removal of non-numeric data.
The latter reports housing data, which, after keeping the 7 dimensions with less nulls and cleaning nulls, consisted of 3,569,678 tuples.

\subsection{Effectiveness of filtering}
\label{sec:effectiveness}

\medskip
\noindent
\textbf{Grid Filtering.}
Grid Filtering exhibits distinct behavior based on the dataset type.
For instance, on uniform datasets with $\dimensions=4$ and sizes varying between $100K$ and $3M$ tuples, 58\% of the tuples are discarded; for correlated datasets, the percentage of discarded tuples attains 90\% on average; with anticorrelated datasets only 16\% of the tuples are filtered.

\begin{figure}%
\centering
\subfloat[][{Uniform}]
{\includegraphics[width=0.3\textwidth]{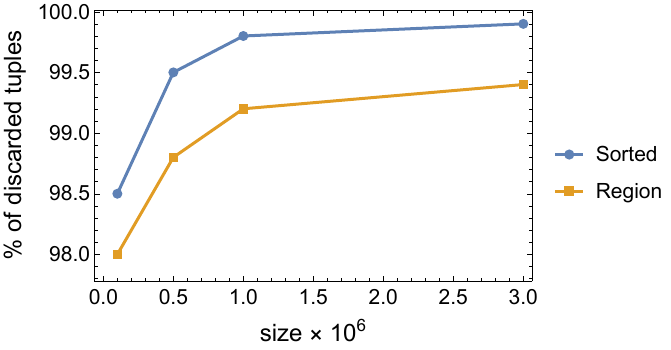}\label{fig:sortedVsRegion-uniform}}%
\quad
\subfloat[][{Correlated}]
{\includegraphics[width=0.3\textwidth]{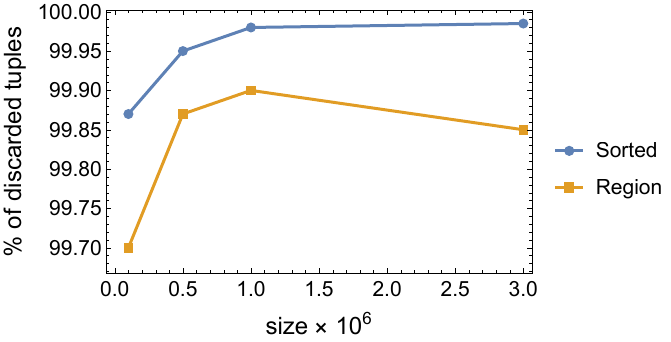}\label{fig:sortedVsRegion-correlated}}%
\quad
\subfloat[][{Anticorrelated}]
{\includegraphics[width=0.3\textwidth]{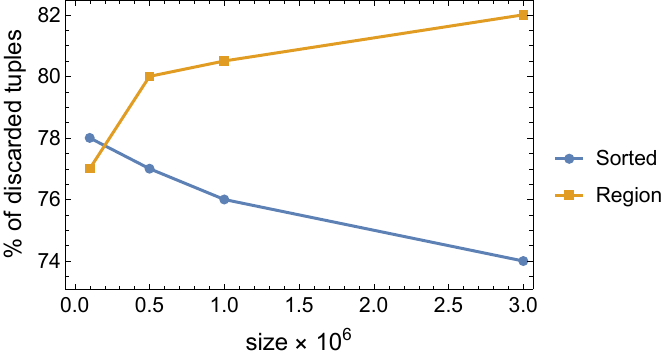}\label{fig:sortedVsRegion-anticorrelated}}%
\caption{Effectiveness of representative filtering: \sortedRep{} and \regionRep{} strategies.}\label{fig:rep-filtering}%
\end{figure}

\medskip
\noindent\textbf{Representative Filtering.}
The percentage of tuples discarded by using representative filtering varies depending on the strategy (\sortedRep or \regionRep) adopted for selecting the representatives and on the type of dataset.
Figure~\ref{fig:rep-filtering} shows the percentage of tuples discarded by filtering with these two strategies on three kinds of synthetic datasets with $\dimensions=4$ dimensions: uniform, correlated and anticorrelated. While \regionRep prevails in the first two cases, \sortedRep is the better option in the more challenging anticorrelated case, and we shall therefore use \sortedRep in the following when adopting Representative Filtering.

\subsection{Efficiency}
\label{sec:efficiency}

In our experiments for testing efficiency, we vary parameters such as dataset cardinality $\size$, number of dimensions $\dimensions$, number of partitions $\partitions$, and number of cores $\cores$ to comprehensively evaluate algorithm performance under diverse conditions.
In experiments where the number of partitions $\partitions$ was kept fixed, for maximum efficiency, we decided to set it equal to the number of cores $\cores$, i.e., $\partitions=120$. However, with \grid or \angular, $\partitions$ cannot be chosen freely, being forced to equal the number of slices $\partitionsPerDim$ raised to the power of $\dimensions$ or $\dimensions-1$, respectively. Therefore, we choose $\partitionsPerDim$ so as to remain as close as possible to the target number -- e.g., $\partitionsPerDim=4$ slices for \grid and $\partitionsPerDim=5$ for \angular with $\dimensions=3$, resulting in $\partitions=256$ and $\partitions=125$ partitions, respectively.

\medskip
\noindent
\textbf{Configuration.}
Our experiments were conducted on a computational infrastructure comprising four virtual machines, each equipped with 30 cores and 8GB of RAM. These machines are interconnected via a Spark cluster, enabling us to leverage the collective computational power of 120 cores and over 30GB of RAM for executing parallel computations.

\begin{table}[h]
   \centering
   \caption{Operating parameters for testing efficiency (defaults in bold).}
       \begin{tabular}{|l|l|}
           \hline
               Full name                           & Tested value \\
           \hline
               Distribution                        & synthetic: \anticorrelated; real: \household, \zillow \\
               Synthetic dataset size ($\size$)                & 100K, 500K, \textbf{1M}, 5M, 10M, 50M, 100M \\
               \# of dimensions ($\dimensions$)    & 2, 3, \textbf{4}, 5, 6, 7 \\
               \# of partitions ($\partitions$)    & 60 $\cdot \{1, \mathbf{2}, 4, 6, 10, 20, 40, 60\}$\\
               \# of cores ($\cores$)    & 4, 16, 48, 80, \textbf{120} \\
           \hline
       \end{tabular}
   \label{tab:operating_parameters}
\end{table}

Besides real datasets, we shall mostly focus on the case of anti-correlated datasets (\anticorrelated{}), which are the most challenging to compute for the skyline operator, since they tend to produce larger sets and, thus, incur more dominance tests. A fuller account on other synthetic dataset types is available in~\cite{pindozzi,delorenzis,pinari}. 
In our analysis, we shall use default parameter values for all the operating parameters described in Table~\ref{tab:operating_parameters}, except for one, which will have varying values, also described in the table.

\begin{figure}%
\centering
\subfloat[][{Global skyline}]
{\includegraphics[width=0.33\textwidth]{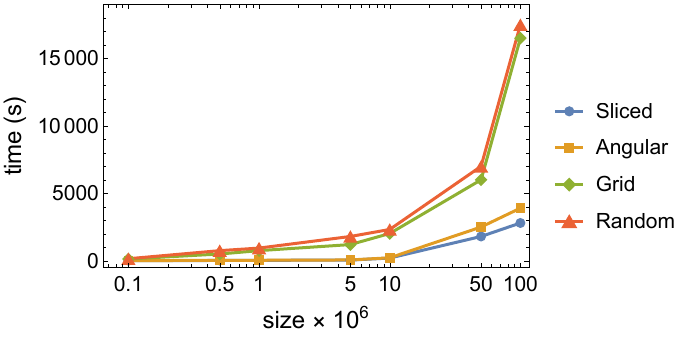}\label{fig:partitioning-global}}%
\subfloat[][{Local skyline}]
{\includegraphics[width=0.33\textwidth]{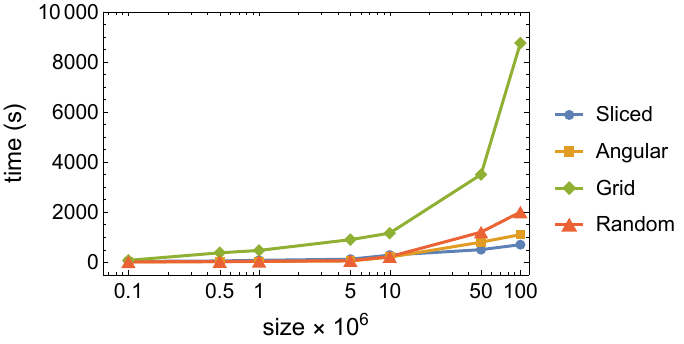}\label{fig:partitioning-local}}%
\subfloat[][{Cardinalities}]
{\includegraphics[width=0.33\textwidth]{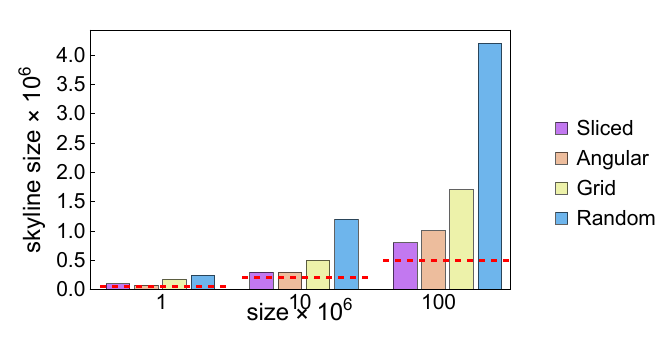}\label{fig:cardinalities}}%
\caption{Performance of different partitioning strategies on an anticorrelated dataset with $\dimensions=4$ dimensions and varying sizes: execution times for computing the global skyline (\ref{fig:partitioning-global}) and the local skylines (\ref{fig:partitioning-local}); numbers of tuples in the local skylines (\ref{fig:cardinalities})}\label{fig:partitioning-varying-size}%
\end{figure}

\medskip
\textbf{Varying size: plain strategies.} We start by considering \anticorrelated with a default number $\dimensions=4$ of dimensions and sizes varying between 100K and 100M tuples on which Algorithm~\ref{alg:parallel-scheme} is applied with the four plain strategies discussed in Section~\ref{sec:parallel}. Figure~\ref{fig:partitioning-global} shows that the overall time to compute the skyline obviously grows as the size grows, with all partitioning strategies, and that the \sliced and \angular strategies perform consistently better than the other two, particularly with larger datasets. A closer look (Figure~\ref{fig:partitioning-local}) reveals that \grid performs poorly in terms of local skyline computation, since much more time is spent in this phase than with the other strategies; instead, Figure~\ref{fig:cardinalities} shows that the baseline \random strategy retains too many tuples in the local skyline computation phase (e.g., $\sim4$M tuples vs $\sim1.5$M with \grid), which will then negatively affect the execution time of global skyline computation, while \sliced and \angular do a much better job.
Overall, \sliced is the best option for this kind of datasets. 

\begin{figure}%
\centering
\subfloat[][{\sliced/\slicedrep}]
{\includegraphics[width=0.33\textwidth]{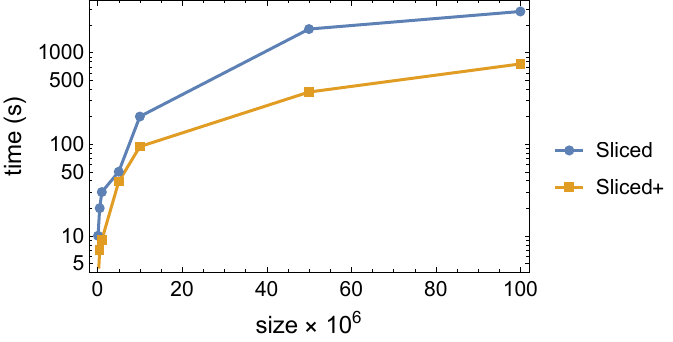}\label{fig:partitioning-improvements-sliced}}%
\subfloat[][{\angular/\angularrep}]
{\includegraphics[width=0.33\textwidth]{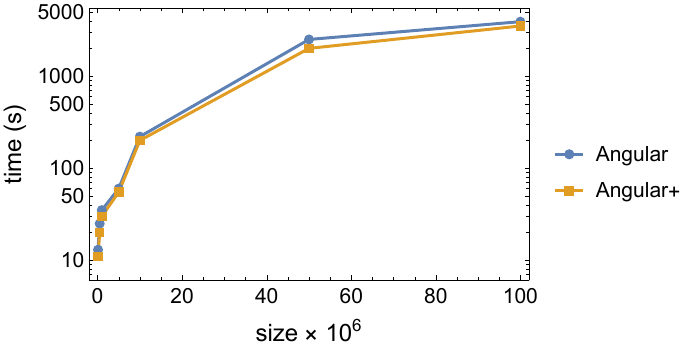}\label{fig:partitioning-improvements-angular}}%
\subfloat[][{All improved algorithms}]
{\includegraphics[width=0.33\textwidth]{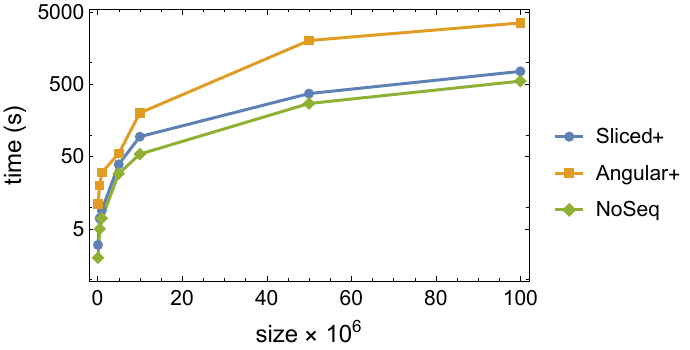}\label{fig:partitioning-all-algos}}%
\caption{Performance of improved partitioning strategies on \anticorrelated and varying sizes: execution times for computing the global skyline with \sliced/\slicedrep (\ref{fig:partitioning-improvements-sliced}), \angular/\angularrep (\ref{fig:partitioning-improvements-angular}) and all improved algorithms (\ref{fig:partitioning-all-algos}).}\label{fig:partitioning-improvements}%
\end{figure}

\medskip
\textbf{Varying size: improved strategies.}
Figure~\ref{fig:partitioning-improvements} shows the effect of applying the improvements discussed in Section~\ref{sec:improvements}. In particular, Representative Filtering is always beneficial in terms of execution times with both \sliced (Figure~\ref{fig:partitioning-improvements-sliced}) and \angular (Figure~\ref{fig:partitioning-improvements-angular}), with larger benefits in the former case (the improved versions are named \slicedrep and \angularrep).
An additional improvement consists in applying the \noseq scheme on top of a partitioning strategy; we choose \sliced for this purpose, which has proved to be consistently better than \angular. As a result, we observe that \noseq prevails over the plain \slicedrep or \angularrep improvements, as shown in Figure~\ref{fig:partitioning-all-algos}.
Considering the degraded performance of \random and \grid, and the fact that the improved versions \angularrep and \slicedrep are consistently better than \angular and \sliced, respectively, we shall henceforth concentrate only on \angularrep, \slicedrep, and \noseq.

\begin{figure}%
\centering
\subfloat[][{\anticorrelated}]
{\includegraphics[width=0.33\textwidth]{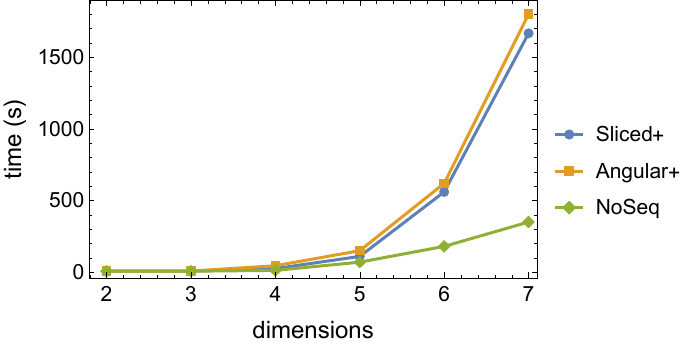}\label{fig:partitioning-dimensions-ant}}%
\subfloat[][{\household}]
{\includegraphics[width=0.33\textwidth]{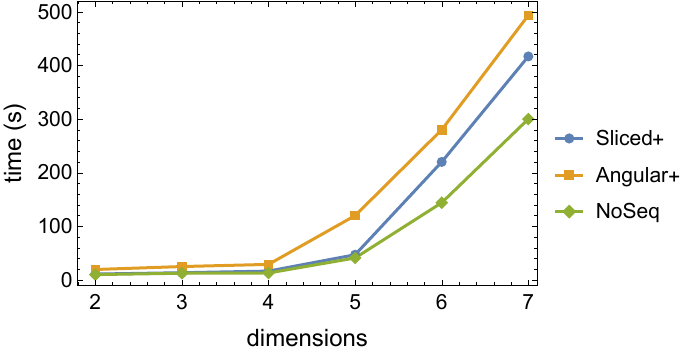}\label{fig:partitioning-dimensions-hou}}%
\subfloat[][{\zillow}]
{\includegraphics[width=0.33\textwidth]{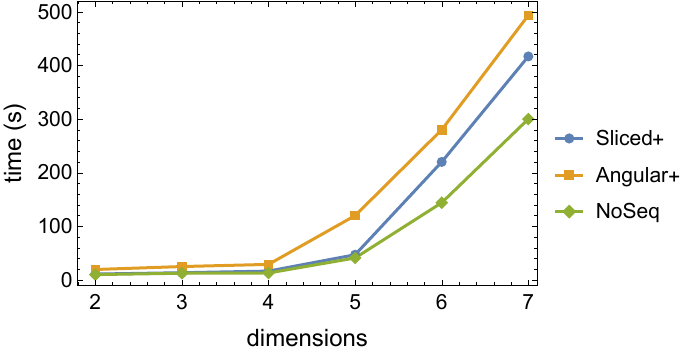}\label{fig:partitioning-dimensions-res}}%
\caption{Execution times for computing the global skyline with improved partitioning strategies varying the number of dimensions on various datasets.}\label{fig:partitioning-dimensions}%
\end{figure}

\begin{figure}%
\centering
\subfloat[][{Varying \# of partitions}]
{\includegraphics[width=0.33\textwidth]{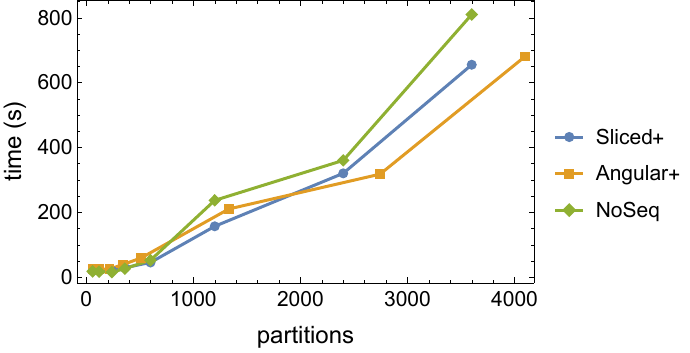}\label{fig:partitioning-partitions}}%
\subfloat[][{Varying \# of cores}]
{\includegraphics[width=0.33\textwidth]{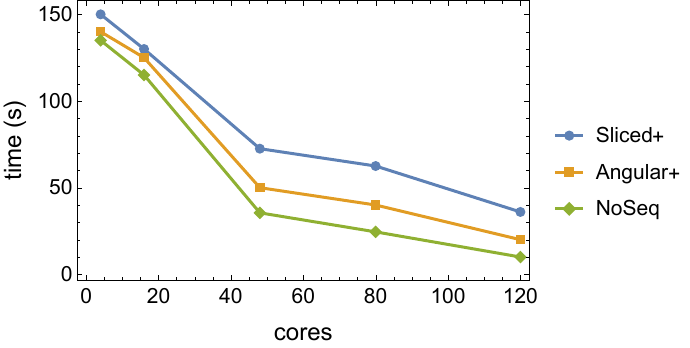}\label{fig:partitioning-cores}}%
\caption{Execution times for computing the global skyline with improved partitioning strategies on \anticorrelated: varying the number of partitions (\ref{fig:partitioning-partitions}) and of cores (\ref{fig:partitioning-cores}).}\label{fig:partitioning-dimensions-partitions-cores}%
\end{figure}

\medskip
\textbf{Varying dimensions.}
Increasing the number of dimensions $\dimensions$ leads to a significant increase in the execution time of various algorithms, which is in turn due to a much increased number of tuples in the skyline. Indeed, when more dimensions are present, it becomes much less likely for a tuple to be dominated, and thus to exit the skyline. Overall, Figure~\ref{fig:partitioning-dimensions-ant} shows that, on \anticorrelated datasets, \noseq remains the most efficient and is the least affected by the curse of dimensionality; this is due to the fact that the size of the union of the local skylines, albeit larger as $\dimensions$ grows, remains manageable, with standard parameter values, so that it can be efficiently passed to all available compute nodes.
Similar observations also hold for the real datasets we considered, however, the relative strength of \angularrep{} and \slicedrep may vary depending on data distribution: in particular, while on the \household dataset (Figure~\ref{fig:partitioning-dimensions-hou}) \slicedrep significantly prevails over \angularrep, consistently with what was observed on synthetic datasets, on the \zillow dataset (Figure~\ref{fig:partitioning-dimensions-res}) \angularrep seems to better capture the data distribution and performs better. A closer inspection reveals that, on \zillow, the size of the local skylines obtained with \angularrep is much smaller than those obtained with \slicedrep, hence the gains.

\medskip
\textbf{Varying number of partitions.}
We observe that all algorithms reach the lowest execution time when around $\partitions=120$ partitions (or at most two or three times as much) are utilized, corresponding to the default number of cores. For larger values of $\partitions$, all algorithms increase their execution time, due to an increased overhead in terms of synchronization between the compute nodes.
Figure~\ref{fig:partitioning-partitions} shows this for several values of $\partitions$ between 60 and 3600 for \slicedrep and \noseq, while for \angularrep the closest values of the form given by Equation~\eqref{eq:index-angular} are used.
This experiment also shows that \noseq is characterized by degraded performance as the number of partitions $\partitions$ becomes too large: indeed, with more partitions, the union of the local skylines grows larger, resulting in more work for the various compute nodes. As an example, with $\partitions=3600$ partitions and standard parameter values, there are 164,183 tuples in the local skyline, while only 27,328 are present with $\partitions=120$.

\medskip
\textbf{Varying number of cores.}
We now examine how the execution times of the algorithms vary when the number of cores $\cores$ varies and all the other parameters use default values. For each machine, we set the number of cores to be one of the following values: 1, 4, 12, 20, and 30; since we use four virtual machines, these values become 4, 16, 48, 80, and 120, respectively.
Clearly, as shown in Figure~\ref{fig:partitioning-cores}, all algorithms benefit from the increased availability of cores, whose utilization is maximized when it coincides with the number of partitions, whose default value was set to $\partitions=120$. We also observe that the largest improvements are visible up to $\cores=48$ cores, while adding further resources after that point produces more moderate benefits.

\section{Related Work}
\label{sec:related}

Since its introduction in the data management community, the skyline operator has stimulated hundreds of research intiatives. Besides the development of efficient algorithms for the centralized case, both sequential~\cite{DBLP:conf/icde/BorzsonyiKS01,DBLP:conf/icde/ChomickiGGL03,DBLP:conf/cikm/BartoliniCP06} and index-based~\cite{DBLP:journals/tods/PapadiasTFS05}, the study of variants (see, e.g.,~\cite{DBLP:journals/tods/CiacciaM20,DBLP:journals/pvldb/CiacciaM17,DBLP:conf/cikm/CiacciaM18,DBLP:conf/sebd/CiacciaM18,DBLP:conf/sisap/BedoCMO19,DBLP:conf/sebd/CiacciaM19,DBLP:conf/sigmod/MouratidisL021} for extensions of the notion of skyline that also include preferences) and the consideration of different architectural solutions have been considered.

For the case in which data are partitioned/distributed, we can distinguish between \emph{vertical partitioning} (in which processing sites/nodes only manage a subset of the attributes involved in the skyline computation) and \emph{horizontal partitioning} (in which each processing node manages a subset of the tuples in the target relation). Solutions for the vertical partitioning case (see, e.g., \cite{DBLP:journals/tkde/TrimponiasBPY13}) are essentially derived from the work of R.\ Fagin on top-k queries and his A0 algorithm~\cite{DBLP:conf/pods/Fagin98}. As already observed in Section~\ref{sec:prelim} and proved in~\cite{DBLP:journals/sigmod/ChomickiCM13}, this follows from the observation that the top-1 tuple for any monotone scoring function is part of the skyline.

The case of horizontal partitioning has been first largely addressed with reference to peer-to-peer (P2P) architectures see, e.g.,~\cite{DBLP:journals/tkde/CuiCXLSX09}. There, the basic idea is to have each peer first compute its local skyline, and then to merge such partial results by injecting them into the network, with aggregation strategies that depend on the specific P2P scenario. For instance,~\cite{DBLP:conf/sebd/BartoliniCP06} proposes to organize the peers' connections (for the purpose of query computation) in a tree-like fashion; then, each peer that is a leaf in such a tree sends the local skyline to its parent peer, which combines these results with its own and recursively propagates until the root of the tree is reached. Interestingly, in order to reduce the amount of transmitted tuples, \cite{DBLP:conf/sebd/BartoliniCP06} proposed to adopt a strategy similar to what here we have called Representative Filtering, yet no actual implementation was demonstrated at that time.

More recently, parallel solutions have been introduced for the Map-Reduce computation paradigm, in which the Mappers distribute data to processing nodes and the Reducer(s) aggregates partial results. Our \noseq optimization has been partially inspired by the technique in~\cite{DBLP:conf/edbt/MullesgaardPLZ14}.

\section{Conclusion}
\label{sec:conclusion}

In this paper, we reviewed the main partitioning methods for computing the skyline in parallel environments (specifically: \random{}, \grid{}, and \angular{}), and proposed a simpler, yet effective method, called \sliced, based on a one-attribute sorting of the dataset.

Given a partitioning method, the typical pattern for computing the skyline consists of two phases: 1) computing, in parallel, a ``local'' skyline for each partition, and 2) eliminating all residual dominated tuples with a final sequential pass computing the skyline of the union of all local skylines.

We introduced two optimization strategies that can be applied on top of this pattern.
The first one, called Representative Filtering, consists in enriching the communication between the nodes with a set of selected tuples with a high potential for dominating other tuples. This strategy proves particularly effective in most experimental scenarios we considered.
The second strategy, called \noseq, aims to eliminate the final sequential phase completely, essentially by passing all the local skylines (or, depending on the partitioning strategy, a selected subset thereof) to all nodes after the first phase, so that the elimination of residual dominated tuples can also be done in parallel. Such a strategy performs very well in datasets with high dimensionality.

According to our extensive experimental evaluation, \sliced and \angular perform consistently better than \grid and \random.
Both optimization strategies are always beneficial, with any number of partitions and cores, but the largest benefits are typically obtained with the \sliced method and the \noseq optimization.
As a final takeaway, the \noseq optimization strategy on top of the \sliced method can be profitably used in all scenarios, unless the number of partitions largely exceeds the available cores (thus causing too large an overhead for \noseq), in which case both \sliced or \angular with Representative Filtering become the preferred alternatives.

Future work will try to adapt these techniques to the computation of skyline variants, typically relying on a modified version of the notion of dominance, which might be exploited for further optimization opportunities in the parallel computation. Another interesting scenario, which would require further investigation, regards the applicability of these partitioning methods for the computation of indicators based, e.g., on the notion of dominance aimed to assess the ``interest'' of a tuple for ranking purposes~\cite{CM:PACMMOD2024}.

\end{document}